# Analytical Research on a Locally Resonant Periodic Foundation for Mitigating Structure-Borne Vibrations from Subway

Yifei Xu[1]; Zhigang Cao[2]; Zonghao Yuan[3]; Yuanqiang Cai, M.ASCE[4]; and Pedro Alves Costa[5]

[1]Ph.D. Candidate, College of Civil Engineering and Architecture, Zhejiang Univ., Hangzhou 310058, China. Email: yifeixu@zju.edu.cn

[2]Professor, Research Center of Coastal and Urban Geotechnical Engineering, Zhejiang Univ., Hangzhou 310058, China (corresponding author). Email: caozhigang2011@zju.edu.cn

[3]Associate Professor, College of Civil Engineering, Zhejiang Univ. of Technology, Hangzhou 310014, China. Email: yuanzh@zju.edu.cn

[4]Professor, College of Civil Engineering, Zhejiang Univ. of Technology, Hangzhou 310014, China. Email: caiyq@zju.edu.cn

[5]Associate Professor, Faculty of Engineering, Univ. of Porto, Porto 4200-465, Portugal. Email: pmbcosta@reit.up.pt

**Abstract:** Filtering properties of locally resonant periodic foundations (LRPFs) have inspired an innovative direction towards the mitigation of structural vibrations. To mitigate the structure-borne vibrations from subways, this study proposes an LRPF equipped with a negative stiffness device connecting the resonator and primary structure. The proposed LRPF can exhibit a quasi-static bandgap covering the ultra-low frequency range. These frequency components have the properties of strong diffraction and low attenuation and contribute the most to the incident wave fields impinging on nearby buildings. By formulating the interaction problem between the tunnel–ground and LRPF–superstructure systems, the mitigation performance of the proposed LRPF is evaluated considering the effects of soil compliance and superstructure. The performance depends on the dynamic properties of the ground, foundation, and superstructure as well as their coupling. Transmission analyses indicate that the superstructure responses can be effectively attenuated in the quasi-static bandgap by adjusting the negative stiffness. Considering the coupling of the flexible ground, the peak responses of the LRPF–superstructure system occur not only at its eigenfrequencies but also at coupled resonance frequencies due to the contribution of the soil compliance. This study provides an analytical tool for mitigating the structure-borne vibrations from subways with the LRPF.

**Keywords:** Locally resonant periodic foundation; Negative stiffness; Bandgap characteristic; Attenuation zone.

# Introduction

Subways have become the efficient transportation infrastructure for alleviating traffic pressure in densely populated modern cities. Frequently operating trains in underground tunnels can generate excessive vibrations and re-radiated noise inside the adjacent buildings, which may cause damages to building structures, breakdowns of sensitive equipment and physical and mental annoyances to residents. The widespread nature of this problem poses a challenge for the research and industry communities that are trying to mitigate the subway-induced structure-borne vibrations and make buildings serviceable to residents.

Numerous measures have been proposed to mitigate the environmental vibrations caused by traffic loads. The most common measures include installing floating slab tracks at the vibration source (Saurenman and Phillips 2006; Hussein and Hunt 2006; Vogiatzis and Kouroussis 2015), using infill or open trenches (Ahmad and Al-Hussaini 1991; Alzawi and EI-Naggar 2011; Cao and Cai 2013; Bose et al. 2018), and installing pile barriers (Kattis et al. 1999; Gao et al. 2006; Cai et al. 2009; Meng and Shi 2018) along the vibration transmission path. The effectiveness of these measures has been analyzed in depth, while these studies have mainly focused on mitigating traffic-induced vibrations at only relatively high frequencies. The frequency components induced by traffic loads can range from quasi-static to much higher frequencies (0-80 Hz) (Lombaert and Degrande 2009). Ultra-low frequency waves dominate the far-field vibrations and are the main contributions to the dynamic responses in the structures adjacent to subways. Few studies till date have focused on mitigating these subway-induced vibrations with ultra-low frequencies.

Periodic structures have drawn extensive attention due to their filtering properties: waves cannot propagate when the frequency of the elastic waves lies in the attenuation zone, the so-called frequency bandgap (Sigalas and Economou 1992; Liu et al. 2000). Therefore, such structures can mitigate vibrations in a unique manner different from the common countermeasures. Based on the formation mechanism of the bandgap, periodic structures are divided into two types: phononic crystals and locally resonant metamaterials (Lu et al. 2009; Hussein et al. 2014). The Bragg scattering bandgap generated by phononic crystals is related to the lattice constant in the periodic structure unit and wave velocity (Bao et al. 2012; Xiang et al. 2012), whereas the local resonance bandgap depends only on the resonant frequency of the resonators in the unit (Sheng et al. 2003, 2007). Thus, locally resonant acoustic metamaterials are more suitable for attenuating low-frequency vibrations than phononic crystals, which require an extremely large lattice constant to generate a low-frequency bandgap (Liu et al. 2005; Wang et al. 2014). To block incident waves from impinging on protected building, two engineering countermeasures based on the concept of locally resonant acoustic metamaterials have been proposed: (1) locally resonant periodic barriers (LRPBs) installed around the buildings (Dertimanis et al. 2016; Lin et al. 2021) and (2) locally resonant periodic foundations (LRPFs) constructed beneath the superstructure (Cheng and Shi 2018; Basone et al. 2019; Xiao et al. 2020). Finocchio et al. (2014) introduced a periodic barrier based on a mass-in-mass chain system to filter the seismic S-waves. The resonance peaks in the soil amplification function could be split into two closer frequencies, enabling the control of soil vibrations. Dertimanis et al. (2016) analyzed the feasibility of using LRPBs for seismic mitigation within an ultra-low frequency bandgap of 0.5-5 Hz. The mitigation performance of two-dimensional and three-dimensional LRPFs was experimentally investigated on scale models (Yan et al. 2014, 2015). Cheng and Shi (2018) proposed a composite periodic foundation consisting of a concrete matrix embedded with steel based on the local resonance concept. The mitigation performance of this periodic foundation in attenuating a set of input ground motions while ignoring the coupling effect of the superstructure was examined. In a similar innovative study, Casablanca et al. (2018) developed a periodic mass-in mass system composed of a concrete plate and steel resonator. Laboratory tests exhibited a frequency bandgap with the lower bound of 4.5 Hz. The above research assumed the LRPF and superstructure as an uncoupled system. To take into account the feedback of the superstructure on the system responses, Bason et al. (2019) and Xiao et al. (2020) analyzed the attenuation behavior of an LRPF subjected to seismic excitation in the LRPF–superstructure system. Frequency shift phenomenon was observed in the superstructure responses due to its interaction with the LRPF, generating additional resonance peaks in the non-attenuation zone and leading to the disappearance of mitigation effects. In addition, some studies demonstrated the promising features of periodic structures equipped with a negative-stiffness element for vibration mitigation. At the material level, Antoniadis et al. (2015) proposed that inserting a negative-stiffness element in the resonator can improve the mitigation performance of a metamaterial-based system. Wenzel et al. (2020) investigated the application of a negative-stiffness element at the structural level. The LRPF implemented a negative-stiffness element was developed for seismic protection of storage tanks. Lin et al. (2021) developed an LRPB with a negative-stiffness element. Its seismic resistance capability was demonstrated through wave-suppression analyses in the ultra-low frequency range. To conclude, a wider bandgap and stronger attenuation can be achieved by implementing a negative-stiffness element. However, existing theoretical and experimental research on periodic structures has primarily focused on seismic protection with little attention paid to the mitigation of subway-induced structure-borne vibrations.

Inspired by these studies, the present study aimed to develop a feasibility analysis method for evaluating the mitigation performance of an LRPF equipped with negative-stiffness elements on subway-induced structure-borne vibrations. In the above-mentioned studies, the LRPFs and LRPBs were usually assumed to be rigidly supported by the ground, and the contributions of the soil compliance were ignored. However, in reality, the contribution of soil compliance can change the boundary conditions of periodic structures, and therefore change the nature of the mitigation phenomena. In addition, the scattering effects due to the presence of the tunnel can also influence soil compliance and eventually change the mitigating properties of the LRPF. Our analysis method can be easily degenerated to the rigid support case, allowing for a better understanding of the interaction effects between the LRPF and ground on the nature of the mitigation phenomena.

Hence, we propose an analytical formulation to evaluate the mitigation performance of the LRPF taking into account the effects of the soil compliance. To this purpose, we calculate the incident wave fields generated by a harmonic source applied at the tunnel invert using the fundamental solution to the tunnel–ground system. This solution is also used as a Green's function to obtain the ground responses generated by the LRPF–superstructure system when subjected to harmonic excitations. The mitigation performance of the proposed LRPF considering soil compliance is then obtained by coupling the tunnel–ground and LRPF–superstructure systems.

The remainder of this paper is organized as follows. The studied problem and the guidelines should be followed are stated first. Subsequently, the fundamental solutions to the tunnel-ground system and the dispersion relations of the LRPF are presented, and the coupling problem between the tunnel–ground and LRPF–superstructure systems is formulated. The next section describes the calculation of the dynamic responses of the LRPF-superstructure system to demonstrate its mitigation performance. The main conclusions of this study are summarized in the closing section.

## Problem Statement

The mitigation performance of an LRPF on the subway-induced structure-borne vibrations is studied. The analytical model is illustrated in Fig. 1. The vibration source is modeled by a vertical harmonic load located at the tunnel invert, which generates incident wave fields in the ground. The building adjacent to the tunnel is excited by the incident wave fields. Fig. 1(b) shows the simplified lumped-mass model for the LRPF and superstructure.

Vibration problems typically involve three aspects: source, propagation and reception, which should be comprehensively considered to develop effective mitigation strategies. For the LRPF to mitigate the subway-induced structure-borne vibrations, the following conditions need to be satisfied. First, the bandgap characteristics of the proposed LRPF must cover the dominant frequencies of the vibration source. Second, the LRPF should be reasonably coupled with ground in the propagation path. To realistically model the foundation–ground interaction, a flexible support provided by the ground is considered to take into account the contributions of the soil compliance instead of assuming the foundation is rigidly supported by the ground. To end with the reception, the coupling effect between the LRPF and superstructure may result in additional resonant peaks, which should be considered when evaluating the actual mitigation performance of the LRPF.

Keeping these guidelines in mind, three fundamental questions need to be considered. The first is to decipher the relationship between the bandgap characteristics of the proposed LRPF and incident wave fields generated by the subway. The second is to understand the influence of soil compliance on the dynamic responses of a superstructure equipped with the LRPF. The third is to reveal the actual performance of the LRPF in mitigating the incident fields taking into account the coupling effect between the LRPF and superstructure.

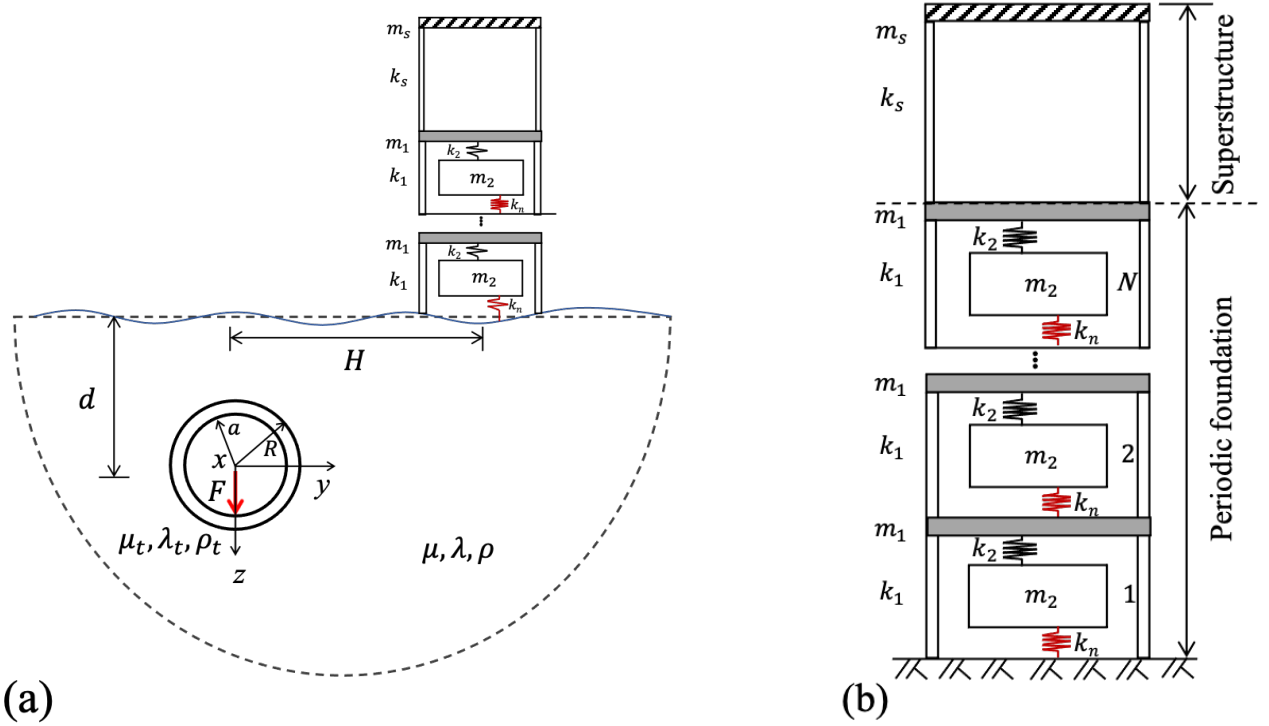

**Fig. 1.** Schematic of (a) tunnel–ground–locally resonant periodic foundation (LRPF)–superstructure model; (b) coupled LRPF–superstructure system.

## Analytical Framework

An analytical framework is established to calculate the dynamic responses of the LRPF-equipped superstructure excited by the incident wave fields from the tunnel.

Our investigation begins recalling fundamental solutions to the tunnel–ground system considering a harmonic point load applied at the tunnel invert or ground surface. Then, we analyze the dynamics and realization of the LRPF. Finally, the responses of the LRPF–superstructure system are obtained by formulating the interaction problem between the tunnel–ground and LRPF–superstructure systems.

### *Fundamental Solution for Tunnel–Ground System*

We consider a harmonic point load applied at the tunnel invert as the vibration source, as shown in Fig. 1(a). The ground is assumed to be made of a linear elastic and homogeneous medium with density $\rho$ and Lamé constants $\mu$ and $\lambda$. Below the ground surface, the tunnel is buried at the center depth $d$. The tunnel wall is modeled as a hollow cylinder with the inner radius $a$ and outer radius $R$, and the material parameters are density $\rho_t$ and the Lamé constants $\mu_t$ and $\lambda_t$.

For the elastic half-space with a cylindrical cavity, the wave function method is used to calculate the wave field, $\mathbf{u}_s$, which is a superposition of the down-going plane waves $U_i^-(q,p,\mathbf{x})$ and outgoing cylindrical waves $U_{im}^o(q,\mathbf{r})$. $q$ and $p$ are the wavenumbers in the $x$ and $y$ directions, respectively. The wave field can be expressed as follows:

$$\mathbf{u}_s = \int_{-\infty}^{\infty}\int_{-\infty}^{\infty}\sum_{i=1}^{3} A_i(q,p)\, U_i^-(q,p,\mathbf{x})\,\mathrm{d}p\,\mathrm{d}q + \int_{-\infty}^{\infty}\sum_{i=1}^{3}\sum_{m=0}^{\infty} B_{im}(q) U_{im}^o(q,\mathbf{r})\,\mathrm{d}q \qquad (1)$$

where $A_i(q,p)$ is the unknown amplitude of the down-going plane wave, which depends on the wavenumbers $q$ and $p$. $B_{im}(q)$ are the unknown amplitude of the outgoing cylindrical wave from the tunnel cavity, which depends only on $q$. $m = 0,1,2,\cdots$. The subscripts $i = 1, 2, 3$ denote the transverse SH wave, transverse SV wave, and longitudinal P waves, respectively. $\mathbf{x}$ and $\mathbf{r}$ are the position coordinates in the rectangular and cylindrical coordinate systems, respectively. The expressions for $U_i^-(q,p,\mathbf{x})$ and $U_{im}^o(q,\mathbf{r})$ are given in Appendices I and II.

The tunnel is modeled as a hollow cylinder and embedded in an elastic half-space. By superposing regular cylindrical waves $U_{im}^{tr}(q,\mathbf{r})$ and outgoing cylindrical waves $U_{im}^{to}(q,\mathbf{r})$, the wave field $\mathbf{u}_t$ in the tunnel wall can be expressed as:

$$\mathbf{u}_t = \int_{-\infty}^{\infty}\sum_{i=1}^{3}\sum_{m=0}^{\infty}\left(C_{im}(q)U_{im}^{tr}(q,\mathbf{r}) + D_{im}(q)U_{im}^{to}(q,\mathbf{r})\right)\mathrm{d}q \qquad (2)$$

where $C_{im}(q)$ and $D_{im}(q)$ are the unknown amplitudes of the regular and outgoing cylindrical waves, respectively, which depend on only $q$. The superscript t denotes the cylindrical waves in the tunnel wall, indicating that tunnel parameters $\rho_t$, $\mu_t$ and $\lambda_t$ should be used. The expressions for $U_{im}^{tr}(q,\mathbf{r})$ and $U_{im}^{to}(q,\mathbf{r})$ are given in Appendix II.

The up-(down-)going plane waves are expressed in rectangular coordinates and the outgoing (regular) cylindrical waves in cylindrical coordinates. Thus, the transformation between plane and cylindrical waves should be adopted to represent these waves at the scattering surfaces in the same coordinates when applying the boundary conditions. The outgoing cylindrical waves from the tunnel cavity propagating in the soil can be transformed into up-going plane waves using the following equation:

$$U_{im}^o(q,\mathbf{r}) = \frac{1}{\pi}\int_{-\infty}^{\infty} U_i^+(q,p,\mathbf{x}) M_{im}(p) \frac{1}{h_i} dp \qquad (3)$$

where $M_{im}(p) = i^{-m}\begin{cases}\sin(m\alpha_i), h_i = h_s, \alpha_i = \alpha_s \ (i=1,2)\\ \cos(m\alpha_i), h_i = h_p, \alpha_i = \alpha_p \ (i=3)\end{cases}$, $\cos\alpha_{s,p} = p/g_{s,p}$, and $h_i$ is the wavenumber in the $z$ direction with non-negative imaginary parts (Yuan et al. 2017a, 2017b). $g_{s,p} = \sqrt{k_{s,p}^2 - q^2}$ are the wavenumbers in the cylindrical coordinate system. $k_{s,p} = \omega/c_{s,p}$ are the wavenumbers in the rectangular coordinate system with transverse and longitudinal wave speeds $c_s = \sqrt{\mu/\rho}$ and $c_p = \sqrt{(\lambda + 2\mu)/\rho}$, respectively. $U_i^+(q,p,\mathbf{x})$ are the up-going plane waves as given in Appendix I.

The down-going plane waves can be transformed to regular cylindrical waves as follows:

$$U_i^-(q,p,\mathbf{x}) = \sum_{m=0}^{\infty} \varepsilon_m U_{im}^r(q,\mathbf{r}) M_{im}(p) \qquad (4)$$

where $\varepsilon_m$ is the Neumann factor defined as $\varepsilon_m = 1\ (m=0)$ and $\varepsilon_m = 2\ (m \gg 1)$. $U_{im}^r(q,\mathbf{r})$ are the regular cylindrical waves from the tunnel cavity.

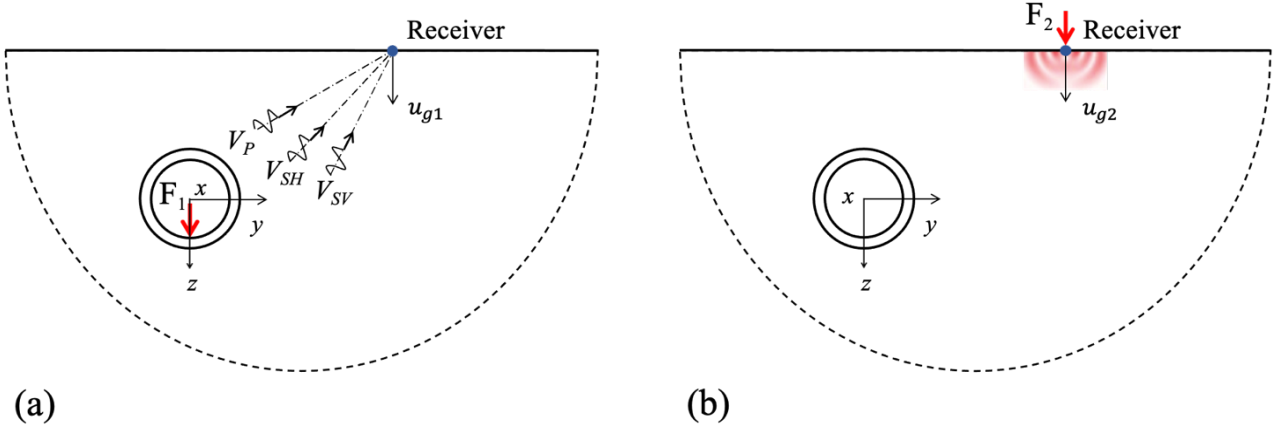

**Fig. 2.** Schematic of two fundamental problems for different vibration sources: (a) external force applied at tunnel invert and (b) external force applied on the ground surface.

The following two sets of boundary conditions are imposed on the four equations to solve these four unknown amplitudes (i.e., $A_i(q, \text{p})$, $B_{im}(q)$, $C_{im}(q)$ and $D_{im}(q)$).

(1) The displacement compatibility and stress equilibrium conditions should be satisfied at the interface of the tunnel and soil ($r = R$).

(2) The external forces applied at the tunnel invert can be considered to obtain the incident wave fields in the ground (see Fig. 2(a)). Specifically, the stress components at the inner tunnel surface ($r = a$) should equal the external forces, and the stress components on the ground surface ($z = -d$) should vanish as the boundary is stress-free. In addition, the external forces applied on the ground surface can be considered to obtain the compliance of the soil in the presence of a tunnel (see Fig. 2(b)). Specifically, the stress components at the inner tunnel surface ($r = a$) should vanish and the stress components on the ground surface ($z = -d$) should equal the external forces.

First, the down-going plane waves in Eq. (1) should be transformed to regular cylindrical waves to apply boundary conditions at the interface of the tunnel and soil ($r = R$), which is expressed as

follows:

$$\sum_{i=1}^{3}\left(\int_{-\infty}^{\infty}\sum_{m=0}^{\infty}\varepsilon_m U_{im}^r(q,r=R)\,A_i(q,p)M_{im}(p)dp+\sum_{m=0}^{\infty}B_{im}(q)U_{im}^o(q,r=R)\right)$$
$$=\sum_{i=1}^{3}\sum_{m=0}^{\infty}\left(C_{im}(q)U_{im}^{tr}(q,r=R)+D_{im}(q)U_{im}^{to}(q,r=R)\right) \quad (5a)$$

$$\sum_{i=1}^{3}\left(\int_{-\infty}^{\infty}\sum_{m=0}^{\infty}\varepsilon_m T_{im}^r(q,r=R)\,A_i(q,p)M_{im}(p)dp+\sum_{m=0}^{\infty}B_{im}(q)T_{im}^o(q,r=R)\right)$$
$$=\sum_{i=1}^{3}\sum_{m=0}^{\infty}\left(C_{im}(q)T_{im}^{tr}(q,r=R)+D_{im}(q)T_{im}^{to}(q,r=R)\right) \quad (5b)$$

where the expressions for $T_{im}^r(q,\mathbf{r})$ and $T_{im}^o(q,\mathbf{r})$ are given in Appendix II.

Then, to obtain the incident wave fields in the ground that can impinge on nearby buildings, an external force $\mathbf{F_1}$ is applied at the tunnel invert. The external force is a harmonic vertical load with a magnitude of 1 N, which has the form of $\mathbf{F_1}=\mathbf{e}_r\delta(\varphi-\pi)\delta(z-ct)e^{-i2\pi f_0 t}$. $c$ and $f_0$ are the speed and excitation frequency of the load, respectively. According to set (2) of boundary conditions, the stress components on the ground surface ($z=-d$) vanish due to the stress-free boundary, and the stress components at the inner tunnel surface ($r=a$) equal the external forces. To apply the boundary conditions on the ground surface, the outgoing cylindrical waves in Eq. (1) should be transformed to up-going plane waves.

$$\sum_{i=1}^{3}\left(A_i(q,p)T_i^-(q,p,z=-d)+\frac{1}{\pi h_i}T_i^+(q,p,z=-d)\sum_{m=0}^{\infty}B_{im}(q)\,M_{im}(p)\right)=\mathbf{0} \quad (6a)$$

$$\sum_{i=1}^{3}\left(C_{im}(q)T_{im}^{tr}(q,r=a)+D_{im}(q)T_{im}^{to}(q,r=a)\right)=\mathbf{e}_r\frac{\varepsilon_m}{4\pi^2 ca}(-1)^m\delta\left(q-\frac{\omega-2\pi f_0}{c}\right) \quad (6b)$$

where the expressions of $T_i^-(q,p,\mathbf{x})$ and $T_i^+(q,p,\mathbf{x})$ are given in Appendix I.

Equations (5) and (6) are a set of linear algebraic equations governing the four unknown amplitudes. Once all unknown coefficients are obtained, the responses of the entire system can be calculated using Eqs. (1) and (2). Thus, the incident wave field in the ground from the tunnel can be obtained, and its vertical component at the receiver is written as $\hat{u}_{g1}$, the superscript '∧' denotes the variable is in the frequency domain.

In addition, to obtain the compliance of the soil in the presence of a tunnel, we apply an external force $\mathbf{F_2}$ exerted at the receiver point on the ground surface. The external force is a harmonic vertical load with a magnitude of 1 N, which has the form of $\mathbf{F_2} = \mathbf{e}_z \delta(y) e^{i2\pi f_0 t}$. According to set (2) of boundary conditions, the stress components on the ground surface ($z = -d$) should equal the external forces, and the stress components at the inner tunnel surface ($r = a$) should vanish.

$$\sum_{i=1}^{3} \left( A_i(q,p) T_i^-(q,p,z=-d) + T_i^+(q,p,z=-d) \sum_{m=0}^{\infty} B_{im}(q) \frac{M_{im}(p)}{\pi h_i} \right) = \frac{\delta(\omega - \omega_0)}{2\pi} \quad (7a)$$

$$\sum_{i=1}^{3} \left( C_{im}(q) T_{im}^{tr}(q, r=a) + D_{im}(q) T_{im}^{to}(q, r=a) \right) = \mathbf{0} \quad (7b)$$

Considering the ground surface boundary conditions, the four unknown amplitudes solved using Eqs. (5) and (7) can be used to calculate the responses of the entire system. The vertical component of the dynamic responses at this receiver, $\hat{u}_{g2}$, can thus be obtained by Eq. (1). Then, the compliance of the soil in the vertical direction induced by the surface force at the receiver can be obtained as

$$H_g(\omega) = \frac{\hat{u}_{g2}}{e^{i2\pi f_0 t}} \quad (8)$$

It should be noted that the compliance of soil $H_g(\omega)$ is an excitation frequency-dependent variable, and its calculation is influenced by the presence of the tunnel.

## *Dynamic Properties of an LRPF*

Now we consider the dynamic properties of an LRPF. The LRPF can be modeled as a simple mass-in-mass lattice system composed of unit cells installed in series, which can suppress wave propagation within a specific frequency range (i.e., bandgap).

To obtain the dispersion relationship of this periodic foundation, the eigenvalue problem of the unit cell needs to be solved by applying the Floquet–Bloch boundary condition. The unit cell is shown in Fig. 3(a), where the outer mass $m_1$ and inner mass $m_2$ are connected by a spring of stiffness $k_2$. Any two adjacent outer masses interact with each other via springs of stiffness $k_1$. A spring of negative stiffness $k_n$ is designed to connect the outer mass of one cell with the inner mass of the immediate next adjacent cell. Fig. 3(b) illustrates an infinite mass-in-mass lattice model.

The equations of motion (EOMs) of this lattice model can be expressed as

$$m_1 \frac{d^2 u_1^j}{dt^2} + k_1\big(2u_1^j - u_1^{j-1} - u_1^{j+1}\big) + k_2\big(u_1^j - u_2^j\big) + k_n\big(u_1^j - u_2^{j+1}\big) = 0 \tag{9a}$$

$$m_2 \frac{d^2 u_2^j}{dt^2} + k_n\big(u_2^j - u_1^{j-1}\big) + k_2\big(u_2^j - u_1^j\big) = 0 \tag{9b}$$

where the superscript $j$ denotes the position of the unit cell, namely, $j$ denotes the unit cell under consideration and $j+1$ denotes the unit cell above it. $u_1$ and $u_2$ are the displacements of the outer and inner masses, respectively.

Based on the Floquet–Bloch theory, the displacements of the masses in the unit cell have the following form:

$$u_1^j(j,t) = \hat{u}_1^j(l,\omega) e^{i(lj-\omega t)} \tag{10a}$$

$$u_2^j(j,t) = \hat{u}_2^j(l,\omega) e^{i(lj-\omega t)} \tag{10b}$$

where $\hat{u}_1^j$ and $\hat{u}_2^j$ are the displacements of the outer and inner masses in the frequency–wavenumber domain, respectively. $t$ and $\omega$ are the time and corresponding frequency, respectively. $j$ and $l$ are the cell position and corresponding wavenumber, respectively. By substituting Eq. (9) into Eq. (10) and eliminating the unknown term $\hat{u}_2^j$, the governing equation involving only $\hat{u}_1^j$ is obtained as follows:

$$\left[m_1 + \frac{m_2(k_2 + k_n)}{k_2 - \omega^2 m_2 + k_n}\right](-\omega^2 \hat{u}_1^j) + \left[k_1 + \frac{k_2 k_n}{k_2 - \omega^2 m_2 + k_n}\right](2 - e^{-iq} - e^{iq})\hat{u}_1^j = 0 \quad (11)$$

Then the relationship between the frequency and wavenumber is as follows:

$$1 - \cos l = \frac{\omega^2 \left[m_1 + \frac{m_2(k_2 + k_n)}{k_2 - \omega^2 m_2 + k_n}\right]}{\left[k_1 + \frac{k_2 k_n}{k_2 - \omega^2 m_2 + k_n}\right]} \quad (12)$$

and the expressions for the effective mass and effective stiffness are

$$m_{eff} = m_1 + \frac{m_2(k_2 + k_n)}{k_2 - \omega^2 m_2 + k_n}; \quad k_{eff} = k_1 + \frac{k_2 k_n}{k_2 - \omega^2 m_2 + k_n} \quad (13)$$

In addition, the characteristic equation can be written as follows:

$$\omega^4 + A\omega^2 + B = 0 \quad (14)$$

where

$$A = -\frac{[2k_1 m_2 (1 - \cos l) + (m_1 + m_2)(k_2 + k_n)]}{m_1 m_2} \quad (15a)$$

$$B = -\frac{2(k_1 k_2 + k_1 k_n + k_1 k_n)(1 - \cos l)}{m_1 m_2} \quad (15b)$$

By varying the wavenumber $l$ from 0 to $\pi$, the lower and upper bounds of the bandgap are obtained as follows:

$$f_L = \frac{1}{2\pi}\sqrt{\frac{1}{2}\left(-A - \sqrt{A^2 - 4B}\right)} \quad (16a)$$

$$f_U = \frac{1}{2\pi}\sqrt{\left(\frac{1}{m_1} + \frac{1}{m_2}\right)(k_2 + k_n)} \quad (16b)$$

where $f_L$ and $f_U$ denote the lower and upper bounds of the bandgap, respectively.

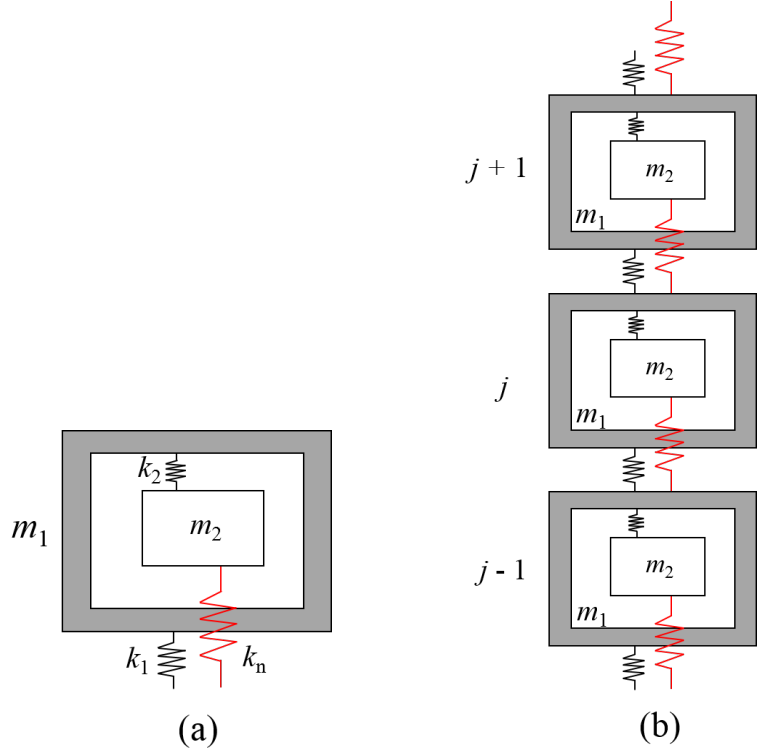

**Fig. 3.** Schematic diagram of mass-in-mass lattice model: (a) basic unit cell and (b) infinite mass-spring lattice.

Before presenting the dispersion relationship, it is helpful to investigate the dimensionality of the problem. From Eq. (12), the lower and upper bounds $f_L$ and $f_U$ can be expressed as functions in the following form:

$$f_L, f_U = \Psi(m_1, k_1, \gamma_m, \gamma_k, \gamma_{kn}) \tag{17}$$

The mass $m_1$ and stiffness $k_1$ of the outer concrete, considering typical material parameter values, are listed in Table 1. The dimensionless parameters $\gamma_m = m_2/m_1$, $\gamma_k = k_2/k_1$ and $\gamma_{kn} = -k_n/k_1$ are introduced to facilitate our investigation. From mathematical considerations, it should be ensured that the lower bound $f_L$ cannot be an imaginary number. When the lower bound $f_L$ equals zero, $\gamma_{kn}$ becomes a critical ratio with the following form:

$$\gamma_{kn} = \gamma_c = \frac{\gamma_k}{1+\gamma_k} \tag{18}$$

The negative stiffness ratio $\gamma_{kn}$ is an important parameter for the bandgap design of a periodic structure. Herein, we first set the mass ratio and stiffness ratio as $\gamma_m = 0.1$ and $\gamma_k = 0.6$, respectively; thus, $\gamma_c = 0.375$. Three negative stiffness ratios $\gamma_{kn} = 0.35, 0.375$ and $0.4$ are considered, resulting in completely different bandgap characteristics, as illustrated in Fig. 4. The lower bound of the bandgap is calculated for $l/\pi = 1$ and the upper bound for $l/\pi = 0$. It can be observed that the lower bound of the bandgap shifts to the quasi-static frequency and a quasi-static bandgap appears when $\gamma_{kn} = \gamma_c = 0.375$. While when $\gamma_{kn} = 0.4 > \gamma_c$, the quasi-static bandgap disappears. In conclusion, the desired quasi-static bandgap can be achieved to cover the quasi-static and ultra-low frequencies by setting $\gamma_{kn} = \gamma_c$, making the LRPF ideal for mitigating subway-induced vibrations where the ultra-low frequency waves dominate in the far field.

Subsequently, if Eq. (18) is always satisfied, we set $\gamma_k = 0.2, 0.6$ and $1.0$. The higher the stiffness ratio $\gamma_k$, the broader the lower bandgap, as shown in Fig. 5. This indicates that the vibration attenuation can be improved by increasing the support strength of the resonators.

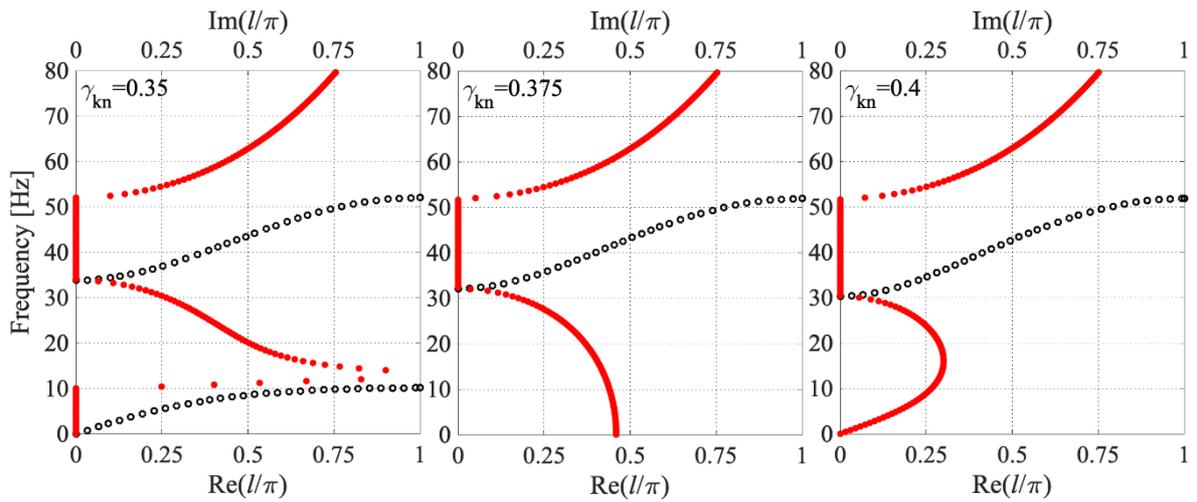

**Fig. 4.** Dispersion curves for varying negative stiffness ratio $\gamma_{kn}$ where the red solid circles and black hollow circles represent the imaginary and real part of wavenumber vector, respectively.

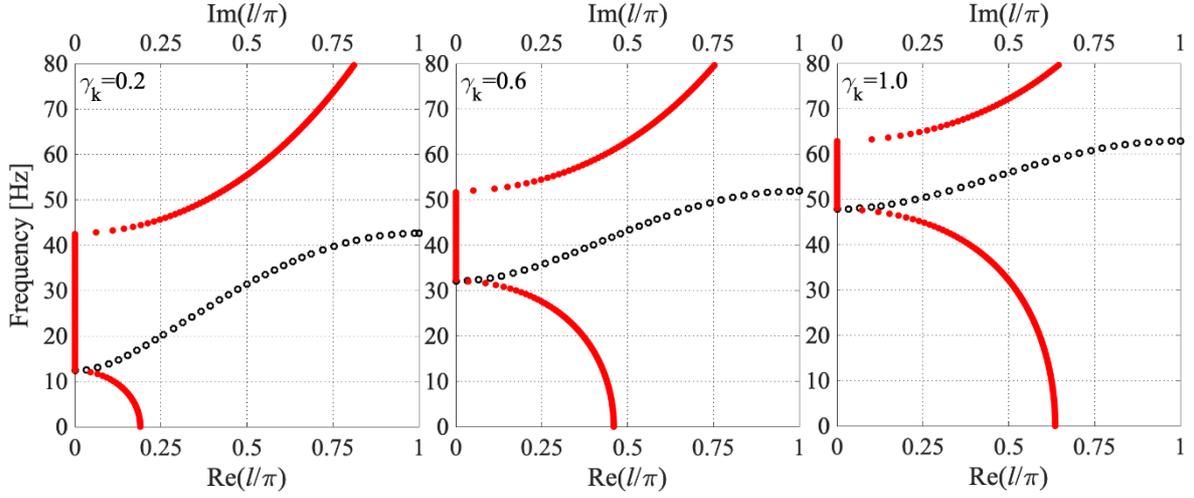

**Fig. 5.** Dispersion curves for varying stiffness ratio $\gamma_k$ where the red solid circles and black hollow circles represent the imaginary and real part of wavenumber vector, respectively.

After presenting the properties of typical unit cell in the periodic structure, in what follows, we show how the negative stiffness element may be realized in the proposed LRPF. Negative stiffness implies that a force is introduced to assist the deformation progress instead of resisting it. In our conceptual design, a compression bar and a preloaded spring are employed to generate the required forces in the vertical general springs, as is schematically illustrated in Fig. 6(a). It can be observed that the compression bar is guided horizontally along the slab and vertically along the common spring such that it can be inclined. When the system deforms in the vertical direction, the precompressed spring can generate a force that is transmitted to the vertical springs by the inclined bar to assist motion, and therefore the purpose of negative stiffness can be achieved. This realization method is inspired by the studies presented by Nagarajaiah and co-workers (Pasala et al. 2013; Sarlis et al. 2013). A simplified analytical model of this practical layout is depicted in Fig. 6(b), where $m_1$ is the mass of concrete slab, $m_2$ the mass of the resonator, $k_1$ the vertical stiffness of all the columns, $k_2$ the stiffness of the spring connecting the slab and resonator, and $k_n$ the stiffness of the spring connecting the resonator with the slab in the neighboring layer.

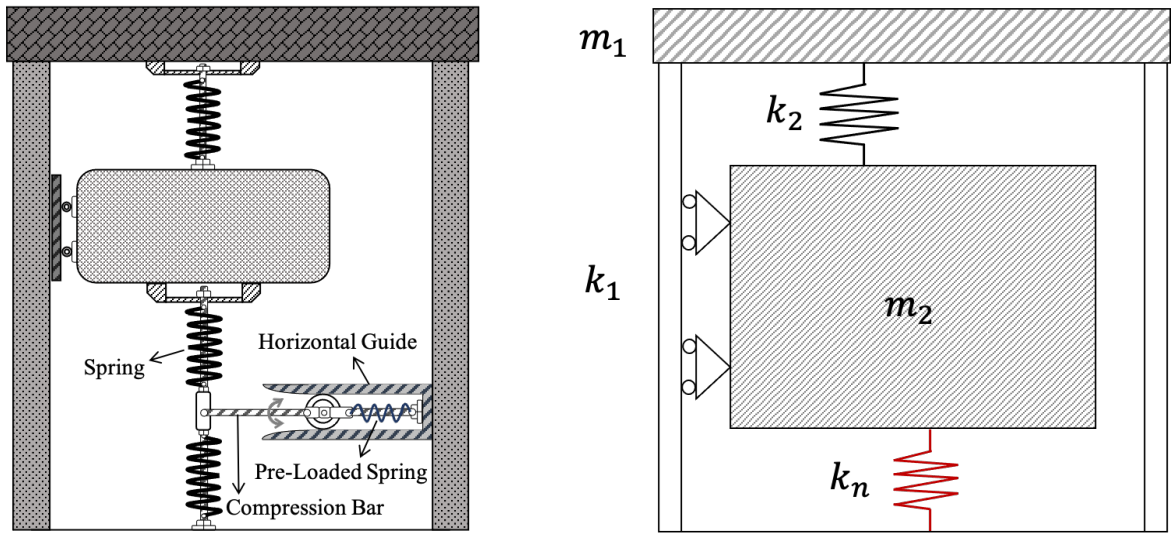

**Fig. 6.** Schematic of (a) realization of negative stiffness spring in LRPF and (b) analytical model of LRPF unit.

## Coupling between LRPF–Superstructure System and Half-space

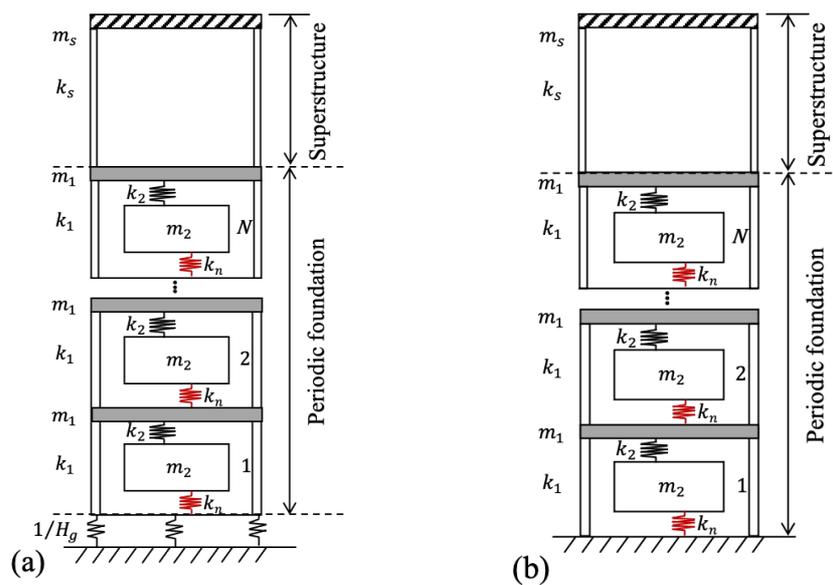

**Fig. 7.** Schematic of *N*-layer LRPF coupled with superstructure assuming (a) flexible ground and (b)

rigid ground.

The equations of motion (EOMs) of the LRPF considering flexible support by the ground (see Fig. 7(a)) are expressed as follows:

$$\mathbf{M}\ddot{\mathbf{u}}^F + \mathbf{K}\mathbf{u}^F = \mathbf{L}F_{fg} \tag{19a}$$

$$\hat{u}_g = H_g(\omega)F_{gf} + \hat{u}_{g1} \tag{19b}$$

where $\mathbf{u}^F(t) = [u_g, u_1^1, u_1^2, \cdots, u_1^N, u_s]^T$, in which $u_g$ is the soil displacement beneath the periodic foundation and equals the displacement at the bottom of the foundation. $u_1^i$ ($i = 1,2\cdots, N$) is the displacement of $i$th layer in the periodic foundation, and $N$ is the layer number. $u_s$ is the displacement of the superstructure. $u_{g1}$ is the vertical component of incident wave field from the tunnel in the absence of the LRPF–superstructure system (i.e., free-field response). $H_g(\omega)$ is the compliance of the soil in the vertical direction induced by the surface force. $u_{g1}$ and $H_g(\omega)$ were obtained in the section *Fundamental Solutions to Tunnel–Ground System*. The mass matrix $\mathbf{M}$ and stiffness matrix $\mathbf{K}$ can be written as follows:

$$\mathbf{M} = \mathrm{diag}(0, m_{eff}, m_{eff}, \cdots, m_{eff}, m_s)_{(N+2)\times(N+2)} \tag{20a}$$

$$\mathbf{K} = \begin{bmatrix} \mathbf{K}_1 & \mathbf{K}_{12} \\ \mathbf{K}_{21} & \mathbf{K}_2 \end{bmatrix}_{(N+2)\times(N+2)} \tag{20b}$$

with

$$\mathbf{K}_1 = k_{eff}; \quad \mathbf{K}_2 = \begin{bmatrix} 2k_{eff} & -k_{eff} & 0 & \cdots & & 0 \\ -k_{eff} & \ddots & \ddots & & & \vdots \\ 0 & \ddots & 2k_{eff} & -k_{eff} & & 0 \\ \vdots & & -k_{eff} & k_{eff}+k_s & -k_s \\ 0 & \cdots & & 0 & -k_s & k_s \end{bmatrix}_{(N+1)\times(N+1)} \tag{21b}$$

$$\mathbf{K}_{12} = [-k_{eff}, 0, \cdots, 0]_{1\times(N+1)}; \quad \mathbf{K}_{21} = \mathbf{K}_{12}^T \tag{21c}$$

where $m_{eff}$ and $k_{eff}$ are the effective mass and effective stiffness, respectively, as given in Eq. (13).

The vector $\mathbf{L} = [1,0,0\cdots,0]^T_{(N+1)\times 1}$ indicates the location of the external excitation applied on the LRPF–superstructure system by the ground. $F_{fg}$ and $F_{fg}$ are the interaction forces; $F_{fg} = -F_{gf} = fe^{i\omega t}$. By substituting Eq. (19b) into Eq. (19a), we obtain the governing equation in the frequency domain:

$$[\mathbf{K} - \omega^2 \mathbf{M}]\hat{\mathbf{u}} = \mathbf{L}\frac{(\hat{u}_g - \hat{u}_{g1})}{H_g(\omega)} \tag{22}$$

The transmission between the LRPF response and the incident wave field can be defined as

$$T = 20\log_{10}\left(\frac{\hat{u}_g, \hat{u}_1^i \text{ or } \hat{u}_s}{\hat{u}_{g1}}\right) \tag{23}$$

By setting the coefficient determinant in Eq. (22) to be equal to zero, the coupled resonance frequencies of the LRPF–superstructure and flexible ground coupled system can be calculated using the characteristic equation. Taking $N = 2$ as an example, the characteristic equation can be expressed as follows:

$$\alpha_1 \omega^6 + \alpha_2 \omega^4 + \alpha_3 \omega^2 + \alpha_4 = 0 \tag{24}$$

where

$$\alpha_1 = m_{eff} m_{eff} m_s \tag{25a}$$

$$\alpha_2 = -\left(2k_{eff} - \frac{k_{eff}^2}{k_{eff} - 1/H_g(\omega)}\right) m_{eff} m_s - (k_{eff} + k_s) m_{eff} m_s - m_{eff}^2 k_s \tag{25b}$$

$$\alpha_3 = \left(2k_{eff} - \frac{k_{eff}^2}{k_{eff} - 1/H_g(\omega)}\right)\left(m_s(k_{eff} + k_s) + m_{eff} k_s\right) + m_{eff} k_{eff} k_s \tag{25c}$$
$$- m_s k_{eff}^2$$

$$\alpha_4 = -\left(2k_{eff} - \frac{k_{eff}^2}{k_{eff} - 1/H_g(\omega)}\right) k_{eff} k_s + k_{eff}^2 k_s \tag{25d}$$

It can be derived that the coupled resonance frequencies are influenced by the soil compliance $H_g(\omega)$, and therefore they are also the frequency-dependent variables.

In addition, Eq. (19) becomes the EOM governing the LRPF–superstructure system coupled with the rigid ground when $H_g(\omega) = 0$. Neglecting the contributions of the soil compliance to the system responses gives the case of rigid support by the ground (see Fig. 7(b)).

**Table 1.** Parameters of tunnel, half-space, and periodic foundation

| Parameter | Value |
|---|---|
| Parameters of tunnel | |
| $a$ (m) | 2.75 |
| $R$ (m) | 3 |
| $\mu_t$ (Pa) | $1.92 \times 10^{10}$ |
| $\lambda_t$ (Pa) | $2.88 \times 10^{10}$ |
| $\rho_t$ (kg/m³) | 2500 |
| | |
| Parameters of half-space | |
| $\mu$ (Pa) | $7.2 \times 10^{7}$ |
| $\lambda$ (Pa) | $1.44 \times 10^{8}$ |
| $\rho$ (kg/m³) | 1800 |
| $d$ (m) | 20 |
| $H$ (m) | 20 |
| | |
| Parameters of periodic foundation | |
| $m_s$ (t) | 700 |
| $m_1$ (t) | 700 |
| $k_1$ (N/m) | $1.15 \times 10^{10}$ |
| $\gamma_m$ (N/m) | 0.1 |
| $\gamma_k$ (N/m) | 0.6 |
| $\gamma_{kn}$ (N/m) | 0.375 |

## Mitigation Performance of LRPF

In this section, the dynamic responses of the superstructure equipped with a two-layer LRPF (i.e., $N = 2$) are investigated to illustrate the mitigation performance of the LRPF, as depicted in Fig. 8. The parameters of the LRPF, ground and tunnel are listed in Table 1. Our objectives are twofold: (1) to demonstrate the mitigation effects of the LRPF in the ultra-low frequency range and (2) to investigate the influence of soil compliance and superstructure on the mitigation performance of the LRPF.

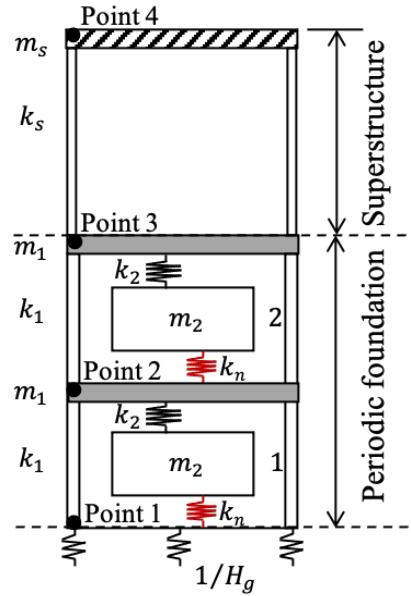

**Fig. 8.** Configuration of the two-layer LRPF and superstructure.

## *Influence of Soil Compliance*

Let us now investigate the relationship between the dynamic responses of the superstructure and soil compliance $H_g(\omega)$, namely support of ground to the LRPF–superstructure system. To this purpose, we first calculate the dynamic responses of an uncontrolled superstructure equipped with a conventional foundation (i.e., without implementing the resonators and negative-stiffness element) on the flexible ground, and compare its responses to the one of an identical structure placed on the rigid ground. The transmission spectra of the two uncontrolled superstructures are plotted in Fig. 9. The responses of the structure lying on the rigid ground are amplified significantly at its eigenfrequencies, whereas for the structure coupled with the flexible ground, the maximum responses do not occur at its eigenfrequencies while shifting to the lower coupled resonance frequencies. The flexible support provided by the elastic ground reduces the magnitude of the transmission spectrum, leading to reduced vibration levels at the resonance peaks. This can be explained by the fact that the contribution of soil compliance is equivalent to providing an additional spring with stiffness $1/H_g(\omega)$, leading to vibration mitigation effects. In addition, in the case of the flexible support, the maximum responses of the ground surface (Point 1) occur at the coupled resonance frequencies, followed immediately by anti-resonances at the eigenfrequencies of the structure, which can be explained by the constraint of structural resonance to the ground.

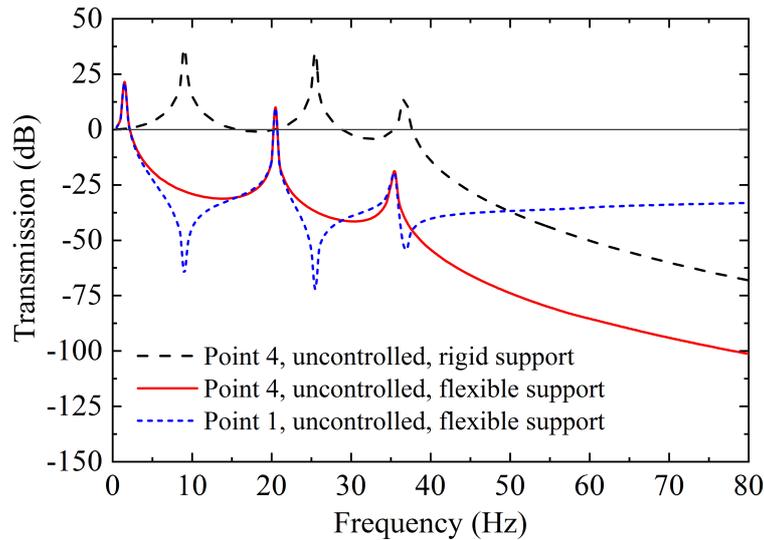

**Fig. 9.** Transmission spectra of uncontrolled superstructure on flexible and rigid ground.

Fig. 10(a) compares the transmission spectra of the superstructure (i.e., at Point 4) controlled by the LRPF lying on the flexible and rigid ground, respectively. The transmission spectra of the uncontrolled superstructure are also presented for comparison purposes. As expected, a significant attenuation zone appears in the quasi-static frequency range starting from zero frequency due to the application of the LRPF. For the case of rigid support by the ground, in which the contribution of soil compliance is not considered, the spectrum of the LRPF–superstructure system is the inherent transmission coefficient of this system. The amplified vibration response of the LRPF–superstructure system (green dashed line) occurs at 23.5 Hz, which is a lower frequency than the second eigenfrequency of the uncontrolled superstructure. In other words, the attenuation zone of the LRPF–superstructure system covers only the first eigenfrequency component of the uncontrolled superstructure. When considering the contribution of the soil compliance, the peak responses of the LRPF–superstructure system (red continuous line) do not only occur at its eigenfrequencies while at coupled resonance frequencies. In other words, in addtion to the three resonance peaks that occur on the green dashed line, two coupled resonance peaks occur on the red continuous line. This can be attributed to the fact that the flexible support by the ground changes the boundary conditions of the LRPF, and therefore the LRPF does not remain a rigorous periodic structure. Moreover, the responses of the LRPF–superstructure system are first amplified at 23.5 Hz, which is a higher frequency than the first two coupled resonance frequencies of the uncontrolled structure (i.e., 1.5 Hz and 20.3 Hz). Thus, the attenuation zone in this case covers the first two coupled resonance frequencies of the uncontrolled structure. Hence, the energy of the ground-borne vibrations can be effectively attenuated, thereby the superstructure responses can be controlled. As shown in Fig. 10(b), by comparing the transmission spectrea of Points 1 and 4, we can find that the transmission between them (i.e., $T = 20\log_{10}(\hat{u}_s/\hat{u}_g)$) is exactly the same as the transmission spectrum of Point 4 in the case of the rigid ground support.

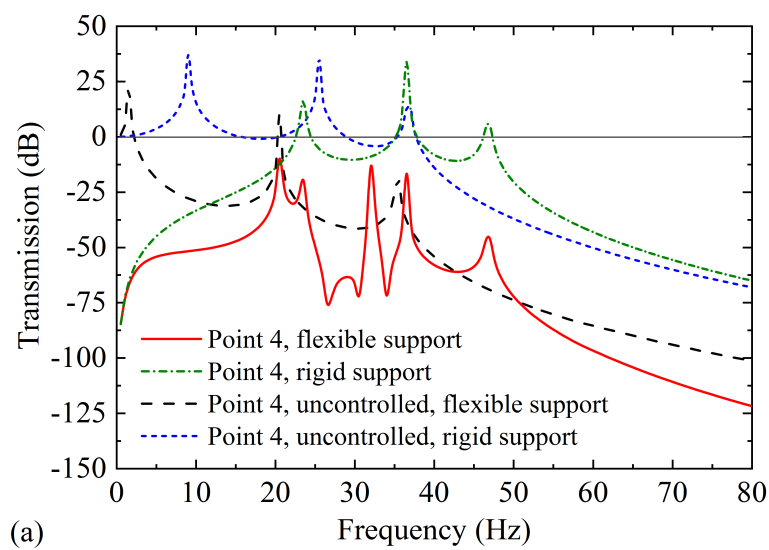

(a)

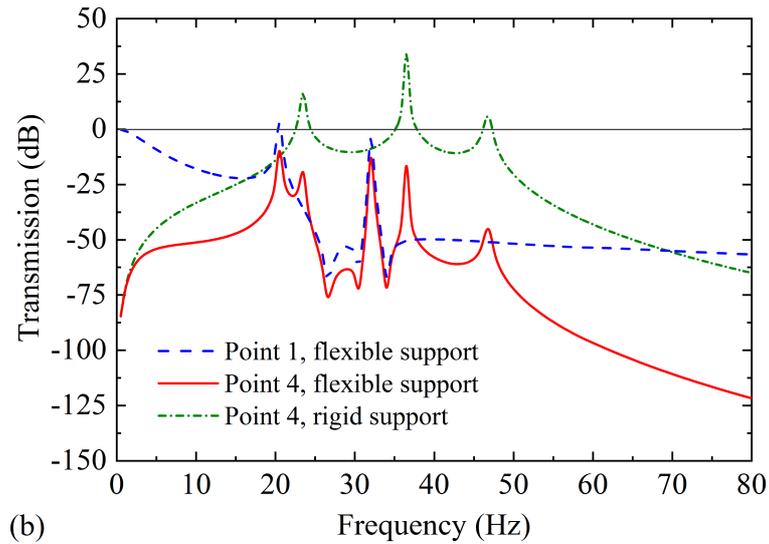

**Fig. 10.** (a) Comparison between the mitigation performances of the LRPF–superstructure system on flexible and rigid ground; (b) relationship between the superstructure response and response at the bottom of the LRPF.

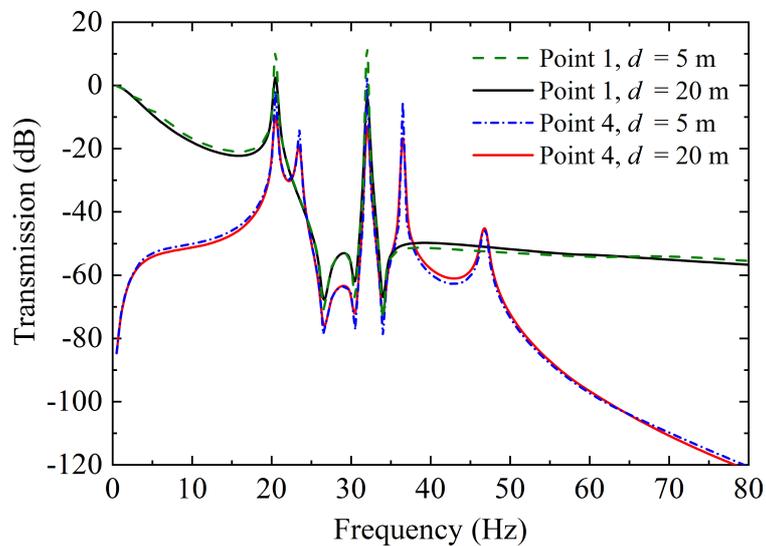

**Fig. 11.** Transmission spectra of the LRPF–superstructure system lying on the flexible ground at different LRPF–tunnel distances.

Note that the calculation of the soil compliance $H_g(\omega)$ is influenced by the presence of the tunnel. Let us then investigate in more detail the relationship between the dynamic responses of the LRPF–superstructure system and $H_g(\omega)$ to highlight the influence of the tunnel. To this purpose, the transmission spectra of the LRPF–superstructure system with different separation distances between the tunnel and LRPF (i.e., $(d = 20\ m,\ H = 0\ m)$ and $(d = 5\ m,\ H = 0\ m)$) are compared in Fig. 11. Different separation distances can cause variations in the $H_g(\omega)$ value due to the varying scattering effects at the tunnel–soil interface. As expected, the transmission spectra of the LRPF are changed by the varying flexible support of the ground. However, the differences between the transmission spectra are not significant. This is likely because the influence of the tunnel can not play a significant role because point contact is assumed between the LRPF and ground, and geometric scattering does not exist.

To conclude, the interaction effects between the tunnel and LRPF–superstructure system need to be considered for the vibration problem induced by subways, which is a significat difference compared with earthquake-related problem. From the above analyses, it can be concluded that the contribution of the soil compliance is important, and therefore it should be considered when applying the LRPF for mitigating the subway-induced structure-borne vibrations.

### *Influence of Superstructure*

The influence of the superstructure on the mitigation performance of the LRPF is evaluated. Fig. 12 displays the transmission spectra of the LRPF–superstructure coupled system at Points 3 and 4, and that of the LRPF at Point 3 without the superstructure. It can be observed that the mitigation performance of the LRPF is reduced when it is coupled with the superstructure. Specifically, the first amplified response at Point 3 of the spectrum of the LRPF without superstructure (green dashed line) occurs at 32 Hz (one coupled resonance frequency), whereas the first amplified response at Point 4 of the LRPF–superstructure system is shifted to 23.5 Hz. Due to the coupling of the superstructure, one observable resonance peak leading to amplified responses appears inside the original attenuation zone, implying that the original attenuation zone is reduced. However, the reduced attenuation zone still covers the first two eigenfrequency components of the uncontrolled superstructure. Thus, it should be pointed out that the mitigation effect is weakened but still satisfactory in the presence of a superstructure. These results demonstrate the necessity of considering the influence of the superstructure in the LRPF–superstructure coupled system as it determines the actual mitigation performance.

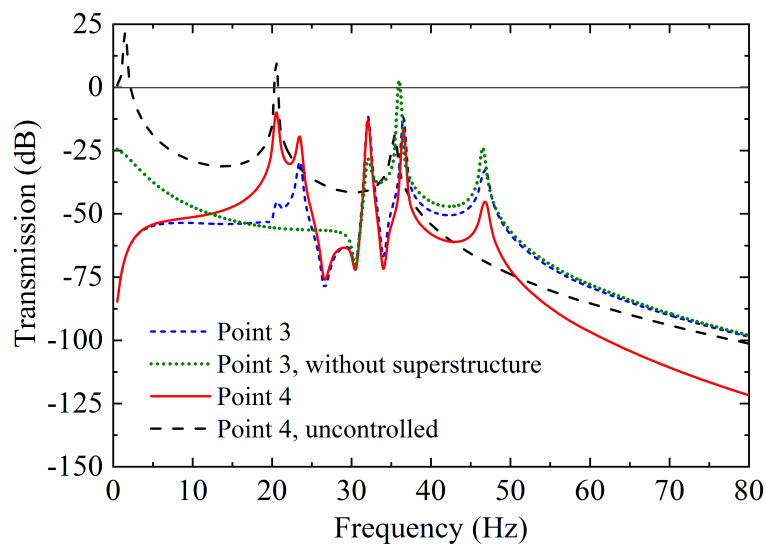

**Fig. 12.** Transmission spectra of the LRPF–superstructure system, LRPF and uncontrolled superstructure on flexible ground.

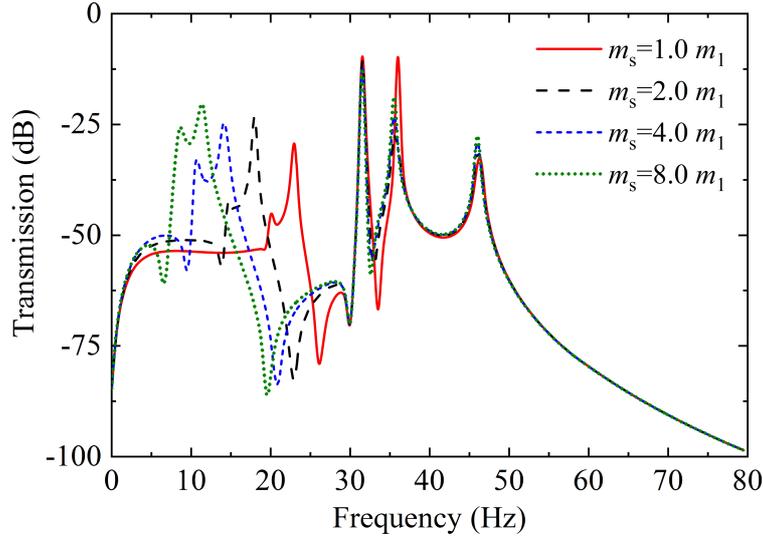

**Fig. 13.** Transmission spectra of the LRPF–superstructure system for varying superstructure mass.

To further reveal the effects of the superstructure on the mitigation performance of the LRPF, we investigate the transmission spectra at Point 3 for varying superstructure mass $m_s$, with results plotted in Fig. 13. It can be seen that this additional resonance peak moves left with the increasing superstructure mass, and thus the attenuation zone is reduced. While the quasi-static bandgap always exists, and therefore the desired mitigation effect can be achieved. The variation in the superstructure mass has negligible effect on the transmission spectra of the LRPF in the relatively high frequency range, where the performance is dominantly influenced by the dynamic properties of the LRPF itself.

*Influence of Negative Stiffness Ratio*

The negative stiffness ratio is another important parameter that determines the mitigation performance of the LRPF. Fig. 14 illustrates the transmission spectra of the LRPF–superstructure system at Point 4 for varying $\gamma_{kn}$. As discussed earlier, setting $\gamma_k = 0.6$ leads to a critical negative stiffness ratio $\gamma_{kn} = 0.375$. The proposed LRPF is transformed to a conventional periodic foundation when $\gamma_{kn} =$

0. When $\gamma_{kn} < \gamma_c$, the central frequency of the bandgap can be observed around the eigenfrequency of resonators, and there are resonance peaks on the left of the attenuation zone. With the increase of $\gamma_{kn}$, the attenuation zone moves toward lower frequencies and attenuation effects gradually occur at the ultra-low frequencies. When $\gamma_{kn}$ reaches the critical ratio, the bandgap displays quite different characteristics, where the attenuation zone starts from zero frequency.

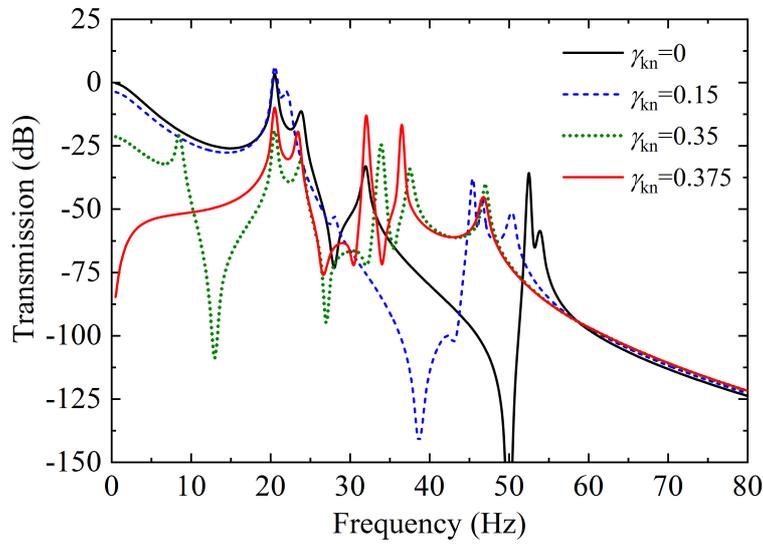

**Fig. 14.** Transmission spectra of the LRPF–superstructure system for varying negative stiffness ratio.

## *Time-History Analysis*

The responses of the LRPF-controlled superstructure are analyzed in the time domain. Fig. 15 shows the displacement time history of the superstructure with and without the LRPF. Here, the velocity of harmonic load moving in the tunnel is 10 m/s, and $t = 0$ s corresponds to the instant at which the moving load and the structure are in the same cross section. When the excitation frequency $f_0$ equals 1.5 Hz falls into the attenuation zone, it can be seen that the peak displacement of the LRPF-controlled superstructure is attenuated by three orders of magnitude compared with that of the uncontrolled superstructure. These results in the time domain further confirm that the proposed LRPF has a prominent ability for mitigating the subway-induced structure-borne vibrations.

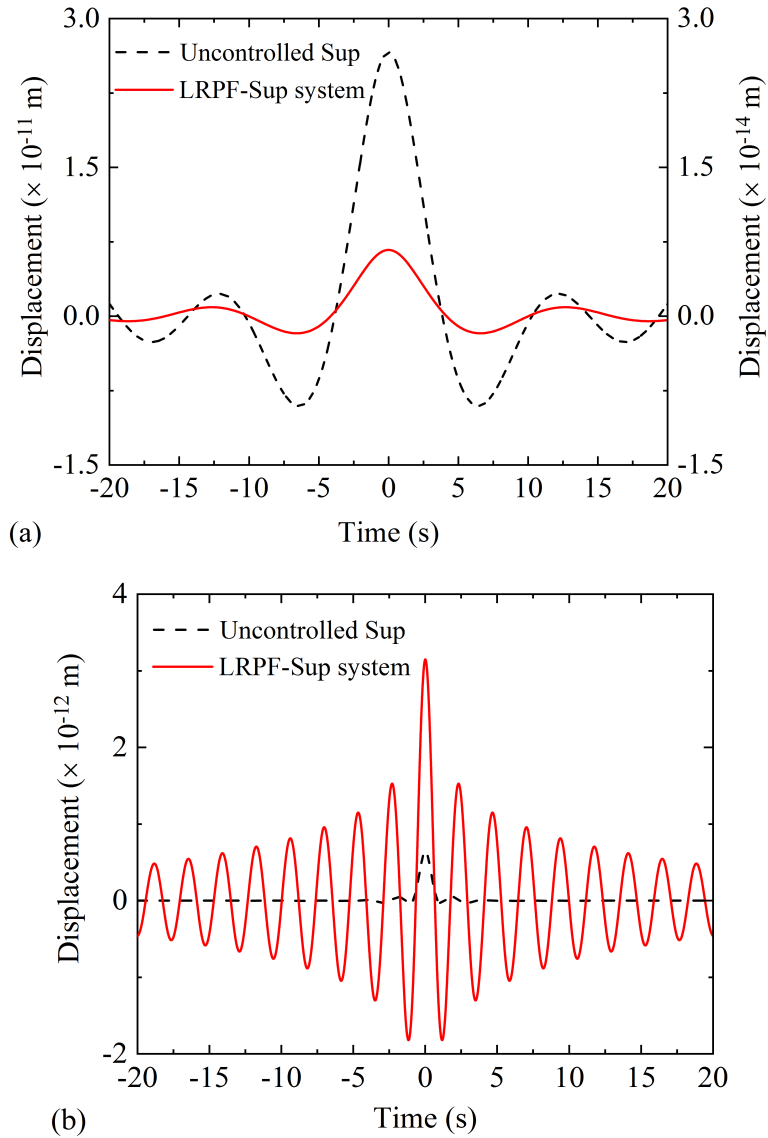

**Fig. 15.** Time history of displacements in superstructure with and without LRPF for different excitation frequencies: (a) 1.5 Hz and (b) 23.5 Hz.

**Conclusions**

In this study, a periodic foundation based on the concept of local resonance is proposed to mitigate subway-induced structure-borne vibrations. By introducing the negative-stiffness vibration absorbers, this locally resonant periodic foundation (LRPF) can achieve a quasi-static bandgap, making it ideal for mitigating subway-induced vibrations where the ultra-low frequency waves dominate in the far field. The mitigation mechanism of the proposed LRPF is investigated in depth, and the influence of the soil compliance and the effects of the superstructure on the mitigation performance of the LRPF are analyzed. Analysis results presented in the frequency and time domains demonstrate its feasibility in mitigating structure-borne vibrations due to subway. The main findings are summarized below.

1. When the negative stiffness ratio $\gamma_{kn}$ is equal to the critical ratio $\gamma_c$, the proposed LRPF exhibits a quasi-static bandgap covering the ultra-low frequency vibrations and main eigenfrequencies of the LRPF–superstructure system, resulting in significant attenuation effects on the structure-borne vibrations.

2. When considering the contribution of the soil compliance, the peak responses of the LRPF–superstructure system coupled with flexible ground occur not only at its eigenfrequencies but also at the coupled resonance frequencies.

3. Additional resonance peaks introduced by the superstructure appear inside the original attenuation zone of the LRPF, resulting in a reduction of the attenuation zone. The presence of superstructure can negatively affect the mitigation performances of the LRPF, thus which should be evaluated in the LRPF–superstructure coupled system.

4. Increasing the stiffness ratio $\gamma_k$ broadens the quasi-static bandgap when the critical negative stiffness ratio is achieved. Increasing the superstructure mass $m_s$ can reduce the attenuation zone of the LRPF–superstructure system.

## Appendix I. Plane Wave Functions

$$U_1^-(q,p,\mathbf{x}) = i(-h_s, -p, 0)e^{i(py-h_s z+qx)} \tag{26a}$$

$$U_2^-(q,p,\mathbf{x}) = (-qp, qh_s, k_s^2 - q^2)e^{i(py-h_s z+qx)} \tag{26b}$$

$$U_3^-(q,p,\mathbf{x}) = i(p, -h_p, q)e^{i(py-h_p z+qx)} \tag{26c}$$

$$T_1^-(q,p,\mathbf{x}) = \mu(p^2 - h_s^2, -2ph_s, qp)e^{i(py-h_s z+qx)} \tag{26d}$$

$$T_2^-(q,p,\mathbf{x}) = \mu i(2qph_s, -2qh_s^2, h_s(2q^2 - k_s^2))e^{i(py-h_s z+qx)} \tag{26e}$$

$$T_3^-(q,p,\mathbf{x}) = \mu(2ph_p, 2k_p^2 - k_s^2 - 2h_p^2, 2qh_p)e^{i(py-h_p z+qx)} \tag{26f}$$

The expressions for $U_i^+(q,p,\mathbf{x})$ and $T_i^+(q,p,\mathbf{x})$ can be obtained by replacing the $-h_{s,p}$ with $h_{s,p}$ in Eq. (26).

## Appendix II. Cylindrical Wave Functions

$$U_{1m}^o(q,\mathbf{r}) = \left(\frac{m}{r}H_m^{(1)}(g_s r)\cos m\varphi, -g_s H_m^{(1)\prime}(g_s r)\sin m\varphi, 0\right)e^{iqx} \tag{27a}$$

$$U_{2m}^o(q,\mathbf{r}) = \left(iqg_s H_m^{(1)\prime}(g_s r)\cos m\varphi, \frac{-imq}{r}H_m^{(1)}(g_s r)\sin m\varphi, g_s^2 H_m^{(1)}(g_s r)\cos m\varphi\right) \cdot e^{iqx} \tag{27b}$$

$$U_{3m}^o(q,\mathbf{r}) = \left(g_p H_m^{(1)\prime}(g_p r)\cos m\varphi, \frac{-m}{r}H_m^{(1)}(g_p r)\sin m\varphi, iqH_m^{(1)}(g_p r)\cos m\varphi\right)e^{iqx} \tag{27c}$$

$$T_{1m}^o(q,\mathbf{r}) = \mu\left(\left(\frac{2mg_s}{r}H_m^{(1)\prime}(g_s r) - \frac{2m}{r^2}H_m^{(1)}(g_s r)\right)\cos m\varphi, -g_s^2\left(2H_m^{(1)\prime\prime}(g_s r)\right.\right.$$
$$\left.\left. + H_m^{(1)}(g_s r)\right)\sin m\varphi, \frac{imq}{r}H_m^{(1)}(g_s r)\cos m\varphi\right)e^{iqx} \tag{27d}$$

$$T_{2m}^o(q,\mathbf{r}) = \mu\left(2iqg_s^2 H_m^{(1)\prime\prime}(g_s r)\cos m\varphi, 2imq\left(\frac{1}{r^2}H_m^{(1)}(g_s r)\right.\right.$$
$$\left.\left. - \frac{g_s}{r}H_m^{(1)\prime}(g_s r)\right)\sin m\varphi, g_s(k_s^2 - 2q^2)H_m^{(1)\prime}(g_s r)\cos m\varphi\right)e^{iqx} \tag{27e}$$

$$T_{3m}^o(q,\mathbf{r}) = \mu\left(\left((k_s^2 - 2q^2)H_m^{(1)}(g_p r) + 2g_p^2 H_m^{(1)''}(g_p r)\right)\cos m\varphi, 2m\left(\frac{1}{r^2}H_m^{(1)}(g_p r)\right.\right.$$
$$\left.\left. - \frac{g_p}{r}H_m^{(1)'}(g_p r)\right)\sin m\varphi, 2\mathrm{i}qg_p H_m^{(1)'}(g_p r)\cos m\varphi\right)e^{\mathrm{i}qx}$$
(27f)

The expressions for $U_{im}^r(q,\mathbf{r})$ and $T_{im}^r(q,\mathbf{r})$ can be obtained by replacing the Hankel function $H_m^{(1)}$ with the Bessel function $J_m$ in Eq. (27). The expressions for $U_{im}^{tr}(q,\mathbf{r})$, $U_{im}^{to}(q,\mathbf{r})$, $T_{im}^{tr}(q,\mathbf{r})$ and $T_{im}^{to}(q,\mathbf{r})$ can be obtained by using the material parameters of the tunnel lining in Eq. (27).

## Data Availability Statement

Some or all data, models, or codes that support the findings of this study are available from the corresponding author upon reasonable request.

## Acknowledgements

The research was supported by the National Natural Science Foundation of China (Grant Nos. 51778571, 51978611 and 52078462), and Natural Science Foundation for Outstanding Scholar in Zhejiang Province (Grant No. LR21E080004) which are gratefully acknowledged.

## Notation

*The following symbols are used in this paper:*

| | |
|---:|:---|
| $a,\ R$ | = inner and outer radii of the tunnel lining, respectively; |
| $A_i(q,\mathrm{p}),\ B_{im}(q)$ | = unknowns in half-space wave field; |
| $C_{im}(q),\ D_{im}(q)$ | = unknowns in tunnel lining wave field; |
| $c$ | = moving speed of the harmonic load; |
| $d$ | = tunnel buried depth; |

| | | |
|---:|:---:|:---|
| $\mathbf{F_1}, \mathbf{F_2}$ | = | external force applied at the tunnel invert and ground surface, respectively; |
| $f_0$ | = | excitation frequency of the harmonic load; |
| $f_L, f_U$ | = | lower and upper bounds of the bandgap, respectively; |
| $F_{fg}, F_{gf}$ | = | interaction forces between the ground and LRPF, respectively; |
| $H_g(\omega)$ | = | soil compliance; |
| $j, l$ | = | location of the unit cell and corresponding wavenumber, respectively; |
| $k_1, k_2, k_n$ | = | spring stiffnesses between outer masses and that between $m_1$ and $m_2$, and negative stiffness between outer and inner mass, respectively; |
| $m_1, m_2, m_s$ | = | masses of outer, inner structure and superstructure, respectively; |
| $m_{eff}, k_{eff}$ | = | effective mass and effective stiffness of the LRPF, respectively; |
| $N$ | = | layer number of the LRPF; |
| $q, p, h_i$ | = | wavenumber in the half-space along *x*, *y* and z directions, respectively; |
| $\mathbf{u}_s, \mathbf{u}_t$ | = | wave field in the half-space and tunnel lining, respectively; |
| $u_1, u_2, u_s$ | = | displacements of outer mass, inner mass and superstructure, respectively; |
| $u_g, u_{g1}, u_{g2}$ | = | vertical response of soil beneath the LRPF, vertical response of the incident wave field and vertical response induced by ground surface load, respectively; |
| $\mathbf{M}, \mathbf{K}$ | = | mass matrix and stiffness matrix of the LRPF–superstructure system, respectively; |
| $\rho, \rho_t$ | = | density of half-space and tunnel lining, respectively; |
| $\mu, \lambda$ | = | Lamé constants of soil; |
| $\mu_t, \lambda_t$ | = | Lamé constants of tunnel; |
| $\gamma_m, \gamma_k, \gamma_{kn}$ | = | dimensionless parameters defining the mass and stiffness, respectively; |
| $\gamma_c$ | = | critical ratio of negative stiffness; |
| $U_i^-(q,p,\mathbf{x}), U_i^+(q,p,\mathbf{x})$ | = | displacement down-going and up-going plane waves in the half-space, respectively; |
| $U_{im}^o(q,\mathbf{r}), U_{im}^r(q,\mathbf{r})$ | = | displacement outgoing and regular cylindrical waves in the half-space, respectively; |

$U_{im}^{to}(q, \mathbf{r}),\ U_{im}^{tr}(q, \mathbf{r})$ = displacement outgoing and regular cylindrical waves in the tunnel lining, respectively;

$T_i^-(q, p, \mathbf{x}), T_i^+(q, p, \mathbf{x})$ = stress down-going and up-going plane waves in the half-space, respectively;

$T_{im}^o(q, \mathbf{r}),\ T_{im}^r(q, \mathbf{r})$ = stress outgoing and regular cylindrical waves in the half-space, respectively;

$T_{im}^{to}(q, \mathbf{r}),\ T_{im}^{tr}(q, \mathbf{r})$ = stress outgoing and regular cylindrical waves in the tunnel lining, respectively;